\documentclass[showpacs,twocolumn,aps,prl]{revtex4}
\usepackage{mathrsfs}
\usepackage{amsmath}
\usepackage{amsfonts}
\usepackage{amssymb}
\usepackage{amsthm}
\usepackage{graphicx}
\usepackage{color}
\usepackage{epsfig}
\usepackage{cancel}

\providecommand{\U}[1]{\protect\rule{.1in}{.1in}}
\usepackage{ulem}
\begin{document}
%----------------------------------------------------------------%
%\title{Level Statistics of Critical States in Two-Dimension}
%\title{Disordered 2DEG at Weak Magnetic Field and Strong Spin-Orbit Coupling}
\title{Absence of Localization in Disordered Two Dimensional Electron Gas at Weak Magnetic Field
and Strong Spin-Orbit Coupling}
\author{Ying Su$^{1,2}$}
\author{C. Wang$^{1,2}$}
\author{Y.  Avishai$^{3,4}$}
\email[corresponding author: ]{yshai@bgumail.bgu.ac.il}
\author{Yigal Meir$^{3}$}
\author{X. R. Wang$^{1,2}$}
\email[corresponding author: ]{phxwan@ust.hk}
\affiliation{$^{1}$Physics Department, The Hong Kong University of
Science and Technology, Clear Water Bay, Kowloon, Hong Kong}
\affiliation{$^{2}$HKUST Shenzhen Research Institute, Shenzhen 518057,
China}
\affiliation{$^{3}$Department of Physics, Ben-Gurion University of
the Negev, Beer-Sheva, Israel}
\affiliation{$^{4}$Department of Physics, NYU-Shanghai University, Shanghai, China}
\date{\today}
%------------------------------------------------------------------%
\begin{abstract}

 The one-parameter scaling theory of localization predicts that all
states in a disordered two-dimensional system with broken time reversal
symmetry are localized even in the presence of strong spin-orbit coupling. While at constant strong magnetic fields  this paradigm fails
(recall quantum Hall effect), it is believed to hold at weak
magnetic fields. Here we explore the nature of quantum states at
weak magnetic field and strongly fluctuating spin-orbit coupling,
employing highly accurate numerical
procedure based on level spacing distribution and transfer matrix technique
combined with finite-size  one-parameter scaling hypothesis.
Remarkably, the metallic phase,
(known to exist at zero magnetic field), persists also
at finite (albeit weak) magnetic fields, and eventually
crosses over into a critical phase, which has already been confirmed
at high magnetic fields.
A schematic phase diagram drawn in the energy-magnetic field plane
elucidates the occurrence of localized, metallic and critical phases.
In addition, it is shown that nearest-level statistics
is determined solely by the symmetry parameter $\beta$ and
follows the Wigner surmise irrespective of whether states are metallic or
critical.
\end{abstract}

\vspace{-0.2in}
\maketitle
%------------------------------------------------------------------%

The one-parameter scaling theory (1PST)  of localization
%The one-parameter scaling theory (OPST) of localization
 \cite{Thouless,Abrahams,Patrick}
has been instrumental in our current understanding of the
% is a powerful tool for studying a possible
metal-insulator transition (MIT) in disordered non-interacting systems. This theory  assumes that the scaling function $\beta(g)$, determining
how the dimensionless conductance $g$ changes with system size, depends
only on $g$ itself, and predicts that the occurrence of a MIT depends
on the symmetry of the system \cite{Wigner,Dyson,Mehta}.
In two dimensions, for both the orthogonal and unitary universality classes under spin rotation symmetry, with and without time reversal symmetry, the 1PST asserts that all states are localized. 
On the other hand, for the symplectic universality class,  where time 
reversal symmetry is maintained but spin rotation symmetry is broken, 
there could be a MIT.  Thus, according to 1PST, despite the presence 
of spin-orbit scattering (SOS), {\it even an infinitesimal magnetic 
field causes all states to be localized}. At high magnetic fields, 
the observation of the quantum Hall effect indicates that extended 
states do exist, but the reason is that in this regime, 1PST should 
be modified to incorporate  two scaling parameters (e.g. the 
conductance and the Hall conductance)  \cite{Pruisken,Huckestein}. 
The question addressed in this work is whether 1PST is still valid (as is 
widely believed) at weak magnetic fields and spatially fluctuating SOS.
Our answer is negative, as  we demonstrate here that under these conditions, 
the band of extended states that exists at zero magnetic field persists 
at weak magnetic fields,  and eventually, with increasing magnetic
field, crosses over into a band of critical states that has been shown to 
exist at strong magnetic fields   \cite{Chen2}.

% In a previous work we have demonstrated that in two-dimensional disordered systems, under strong magnetic field and strongly fluctuating SOC there is a band of critical states. The natural question is then: how does the band of extended metallic states (EMS)  \cite{note0} that exists at zero magnetic field and strong SOC develop into the band of extended critical states (ECS) at strong fields? Recall that OPST predicts the absence of extended states at finite (no matter how small) magnetic fields.

In order to substantiate our claim, we study the nature of non-interacting 
electronic states in 2D under the influence of weak magnetic field,
disorder potential and strongly fluctuating SOS, and carry out two kinds 
of numerical calculations:
The first one studies the nearest level spacing distribution in various  
energy regimes, in order to identify the localized phase and the 
appropriate universality classes \cite{Wigner,Dyson,Mehta}.
The second one consists of highly accurate procedure for identifying MIT, 
based on the transfer matrix technique and finite-size scaling arguments. 

In weak magnetic fields, the Landau levels mix and projection on the 
lowest Landau level is meaningless.  An appropriate and convenient procedure
is then to consider a  tight-binding model for
2D electrons hopping on a square lattice of unit lattice constant.
The lattice sites are labeled as $i=(n_i,m_i)$, with
$1 \le n_i \le L$ and $1 \le m_i \le M$ integers. The Hamiltonian reads,
\begin{equation}
\begin{gathered}
{H}=\sum_{i,\sigma}\epsilon_{i}c^{\dagger}_{i,\sigma}c_{i,\sigma}
+\sum_{\langle ij\rangle,\sigma,\sigma'}\exp(i\phi_{ij})
V_{ij}(\sigma,\sigma')c^{\dagger}_{i,\sigma}c_{j,\sigma'}.
\end{gathered}
\label{2DSU}
\end{equation}
Here
$c^{\dagger}_{i,\sigma}$ ($c_{i,\sigma}$) is the electron creation
(annihilation) operator at site $i$ with spin projection $\sigma=\pm$,
and $\langle ij\rangle$ denotes nearest-neighbour lattice sites. The on-site energies $\epsilon_i$ are randomly distributed in
$[-W/2,W/2]$, (hereafter we take $W=1$  \cite{note1}), and the magnetic
field is introduced by the Peierls substitution in which phase factors
$\phi_{ij}=(e/\hbar)\int^{j}_{i}\vec{A}\cdot d\vec{l}$ multiply the
hopping amplitudes, where $\vec{A}$ is the vector potential  \cite{xrw1}.
The dimensionless parameter $B$ is defined such that magnetic flux
through a unit cell is $ B\times \phi_0$ where $\phi_0\equiv hc/e$ is
the quantum flux unit. Accordingly, $B$ is a measure of
the magnetic field strength in this lattice model. \\
 \ \\
 %we define the strength of the magnetic field $B\equiv\phi/\phi_0$.
 
The SOS is encoded
by random $SU(2)$ matrices $V_{ij}$ acting in spin space, defined as,
\begin{equation}
\begin{gathered}
V_{ij}=
\left(\begin{array}{cc}
e^{-i\alpha_{ij}}\cos(\beta_{ij}/2) &
e^{-i\gamma_{ij}}\sin(\beta_{ij}/2) \\
-e^{i\gamma_{ij}}\sin(\beta_{ij}/2) &
e^{i\alpha_{ij}}\cos(\beta_{ij}/2)
\end{array}\right)
\end{gathered}
\label{SOC}
\end{equation}
where $\alpha_{ij}$ and $\gamma_{ij}$ are uniformly and independently
distributed in a range $[0,2\pi]$, while $\cos\beta_{ij}$ is uniformly
distributed in $[0,1]$. This model is hereafter referred to as the 2DSU model.
For $B=0$ it displays the (so called) symplectic MIT, pertaining to 
systems with conserved time reversal and broken spin rotation symmetry,
as also predicted within 1PST \cite{Hikami}. For strong magnetic field 
(e.g. $B \ge 1/5$) the 2DSU model exhibits a Berezinskii-Kosterlitz-Thouless transition (BKTT) between a band of 
localized states and a band of critical states  \cite{Chen2}.
In the following we concentrate on the physics at weak magnetic fields, 
and even reach $B< 10^{-4}$.

We first concentrate on the distribution $P(s)$ of nearest level spacings 
$s$ (in units of the mean level spacing). This analysis enables the 
distinction between localized and extended states, and in the latter case,  
identification of the relevant universality class: More concretely,
for {\it localized states}, it is expected to follow the Poisson 
distribution  $P_{\mathrm{Loc}}(s)=\exp[-s]$, while for {\it extended 
states},
%{\YA the variance $\Sigma_2(N)$ of
%the number $N$ of energy levels in a given energy interval, and the
%multi-fractal dimensions $f(\alpha)$ (measure the multi-fractal structure
%of the wave function),} are adequately described by the formalism
%of random matrix theory (RMT) \cite{Wigner,Dyson,Mehta,Mirlin}.
$P_\beta(s)$ is specified by the symmetry parameter $\beta=1,2,4$
%given symmetries under time reversal and spin rotation symmetries are
(corresponding respectively to the Gaussian orthogonal (GOE), unitary (GUE)
and symplectic (GSE) ensembles). These three distributions
are excellently approximated by the Wigner surmise expressions $P_\beta (s)=C_1(\beta) s^\beta\exp[-C_2(\beta)s^2]$. (The constants $C_1$ and $C_2$ are determined by normalization conditions for probability and
unit mean level-spacing $\langle 1 \rangle=\langle s \rangle=1$).

\begin{figure}
  \begin{center}% Requires \usepackage{graphicx}
  \includegraphics[width=8 cm]{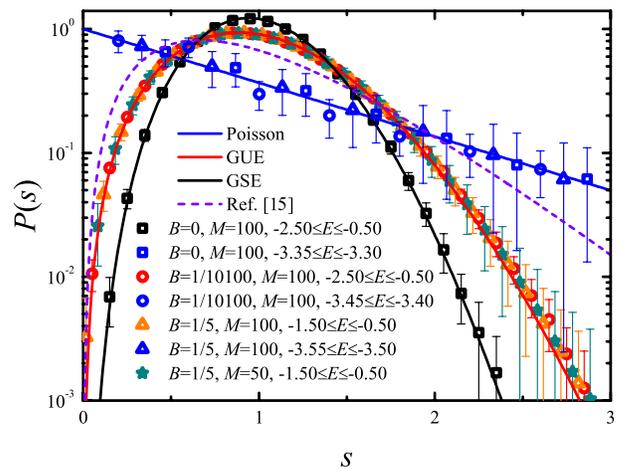}
  \end{center}
  \vspace{-0.2in}
\caption{(color online). \footnotesize{%Nearest level spacing distribution
$P(s)$ for $W=1$, various magnetic field strengths $B=$0, 1/10100,  1/5,
various system sizes $M=$50, 100, and in various energy ranges  \cite{note3}.
%as denoted by the various symbols in the figure.
Data are averaged over 1500 ensembles.
% for $B=$0, 1/10100 and 1500 ensembles for $ B=$1/5.
It is evident that for $B>0$ (no matter how small), $P(s)$ corresponding
to {\it both critical and extended states} fits well into the Wigner surmise for GUE (red solid line), whereas $P(s)$ corresponding to
extended states at $B=0$ agree with the Wigner surmise for
GSE (black solid line). The dashed line corresponds to the distribution
 suggested in  Ref.  \cite{newclass} assuming $\nu=\infty \Rightarrow \gamma=1$
since the localization length at a BKTT diverges faster than a power-law.
%(near the mobility edge) faster than a power-law.
For localized states with energies $-3.55\leq E \leq -3.50$
far from BKTT ($E_c\simeq -3.0$ for $B=1/5$), whose localization length
is much smaller than the sample size ($M=100$) $P(s)$ agrees with the
Poisson distribution (blue solid line).
%%%The dash line is Eq.~\ref{ps}.
}}
  \label{fig1}
  \vspace{-0.2in}
\end{figure}

For the actual computation, a finite lattice of size $M\times (M+1)$ is 
considered and periodic boundary conditions are imposed on both directions 
using the almost antisymmetric gauge suggested in Ref.~ \cite{note2}.
That makes it possible to treat a weak field $B=\frac{1}{M(M+1)}$.
The Hamiltonian (\ref{2DSU}) is diagonalized, yielding all eigenvalues 
$\{ E \}$ and normalized
wave functions $\{ \psi_E(n_i,m_i) \}$ for each  value of $B$ and $M$.
As shown in Fig.~\ref{fig1}, $P(s)$ for B=0 and strong SOS displays, for a 
wide energy range $-2.5\leq E \leq -0.5$, GSE statistics (data in black 
squares, theory in black curve). It suggests the existence of a band of 
extended states within the symplectic ensemble, commensurate  with the 
prediction of 1PST \cite{Hikami}. Remarkably,  {\it adding a single flux 
through the entire area}, corresponding to $B=1/10100$ for $M=100$ (red 
circles) is already sufficient to modify $P_{\beta=4}(s)$ into $P_{\beta=2}(s)$,
where the level statistics follows the GUE Wigner surmise (red line in 
Fig.~\ref{fig1}). In any case, the fact that in both cases $P(s)$ follows the Wigner surmise and not Poisson distribution indicates that these are 
{\it metallic-like states, where level repulsion occurs at small $s$. }
This behavior persists for different system sizes and for all $B>0$.
On the other hand, for energies below the mobility edge (blue shapes and 
curve in Fig.~\ref{fig1}),
$P(s)$ obeys Poisson statistics, as expected for localized states.
Thus,  our analysis of nearest level spacing distribution
suggests that states in the same energy range (as for $B=0$)
are still extended at finite magnetic field even though this 2D system 
now belongs to the unitary class.
 %This result is incommensurate with the predictions of 1PST.
%\YM{I would like to see Poisson statistics also for $B=0, 1/10100$.}
%{\YA I understand that that is now shown in the figure (blue circles with holes.}
The wide range of parameters and energies, where the GUE statistics has been observed, indicates this result is robust and is not due to finite size effects. Further substantiation of this statement is presented below.

\begin{figure}
  \begin{center}% Requires \usepackage{graphicx}
  \includegraphics[width=8 cm]{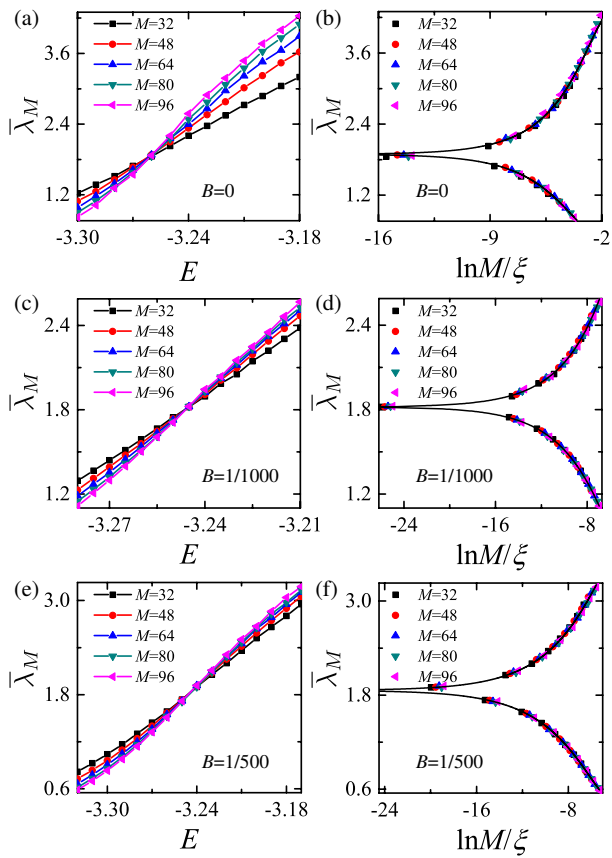}
  \end{center}
  \vspace{-0.2in}
  \caption{(color online). \footnotesize{The left panel displays
  $\bar\lambda_M\equiv\lambda_M/M$ {\it v.s}  $E$
calculated for disorder strength $W=1$ and for three values of the magnetic field
(a) $B=0$, (c) $B=1/1000$, and (e) $B=1/500$ for $M=$32 (squares),
48 (circles), 64 (up-triangles), 80 (down-triangles), and 96 (left-triangles). The scaling function obtained from (a), (c), and (e) by collapsing data of $\bar\lambda_M$ near the transition points into a single curve $\xi\sim(E-E_c)^{-\nu}$ are shown in (b) for $B=0$, (d) for $B=1/1000$, and (f) for $B=1/500$.
}}
  \label{fig2}
 \vspace{-0.2in}
\end{figure}

In order to corroborate our finding on the existence of extended states at weak 
magnetic field (that is so far based on level spacing analysis of finite size systems),
we directly evaluate the localization length $\xi(E,B)$ of the 2D system (up to a 
multiplicative constant) employing the transfer matrix technique \cite{localization}.  
Within this procedure, one evaluates the localization length $\lambda_M$ 
of a stripe of width $M$ and (virtually infinite) length $L>10^6$.
According to the  scaling analysis, the renormalized localization length of the 
strip, $\bar\lambda_M\equiv \lambda_M/M$, increases (decreases) with $M$ for 
extended (localized) states and is independent of $M$ for critical states.  
For the 2DSU model, Figs.~\ref{fig2}(a,c,e) display  $\bar\lambda_M$ {\it v.s} 
$E$ for $B=0$, $B=1/1000$, and $B=1/500$. It is clear from these figures that the system 
undergoes an Anderson MIT, since all curves for different $M$ cross at two 
mobility edges at which $d \bar\lambda_M/d M$ changes sign.
The results of Fig.~\ref{fig2}(a)  just reconfirm the familiar symplectic MIT, 
but the MIT displayed in Fig.~\ref{fig2}(c,e) occurring at mobility edges 
$E_c=\pm 3.245$ and $\pm 3.242$ is novel, and agrees with the conclusion based on level-spacing 
analysis: In the presence of strong SOS fluctuations, a band of extended states 
occurs in 2D systems  even when its Hamiltonian breaks time-reversal symmetry.

%%%%%%%%%--------Collapse figure here, Fig 2----------------
%\begin{figure}
%  \begin{center}% Requires \usepackage{graphicx}
%  \includegraphics[width=8 cm]{Fig2.eps}
%  \end{center}
%  \vspace{-0.2in}
%  \caption{(color online). \footnotesize{The upper panel displays
%  $\bar\lambda_M\equiv\lambda_M/M$ {\it v.s}  $E$
%calculated for disorder strength $W=1$ and for two values of the magnetic field
%(a) $B=1/1000$ and (b) $B=1/10$ for $M=$32 (squares),
%48 (circles), 64 (up-triangles), 80 (down-triangles), and 96 (left-triangles).
%(c) Schematic phase diagram in the $B-E$ plane
%displaying the occurrence of three phases of states, localized (pink),
%EMS (blue) and ECS (green). See text for further details.
%{\color {red} Yshai: The caption does not match the figure!!}}}
%  \label{fig2}
% \vspace{-0.2in}
%\end{figure}

To  substantiate that these results are not merely due to finite size effects, 
we employ the one parameter finite-size scaling formalism, which is based on 
the hypothesis $\bar {\lambda}_M=f(x)$, where $x=M/\xi=C M/(E-E_c)^{-\nu}$. 
Here $C$ is a constant and $\nu$ is the localization-length critical exponent.
For optimal values of $E_c$ and $\nu$, the scaling function $f(x)$ should be 
smooth (actually there are two functions, one for the insulator and one for the 
metallic side). The numerical values of $\nu$ %and $f(0)$
characterize the 
universality class of the MIT \cite{Obuse}.
In Fig. \ref{fig2}(b) the different curves of Fig.~\ref{fig2}(a), when plotted
as function of $x$, indeed collapse on a smooth curve that represents the scaling
function $f(x)$. Here, for $B=0$, this result reconfirms the criticality of
the symplectic MIT. The value of $\nu$ %and $f(0)$
(see first row of the table below)
agrees with previous ones \cite{Ohtsuki,Obuse,Janssen}. Remarkably, 
inspection of Fig.~\ref{fig2}(d,f) shows that the collapse scenario occurs also
{\it at finite magnetic field}, namely the different curves in
Fig.~\ref{fig2}(c,e) fall on a single smooth curve. Moreover, for these novel MIT at $B>0$, the
dependence of $\nu(B)$ %and $f(0,B)$ 
on $B$ is dramatic
and even puzzling (see table). 
This gradual increase of $\nu$ is most likely due to the transition from Anderson MIT to BKKT (where, by definition, $\nu\rightarrow\infty$, that occurs around $B=1/70$. 
%In the thermodynamic
%limit, we expect $\nu$ to be a universal constant, instead of vary with $B$. \\
%{\YMq{I worry about this point, we have claimed so far that there are no finite size effects, but now we do ?} \\
Thus, the two analyses confirm the existence of extended states
 for $0 \le B \lesssim 1/70$. 

\begin{table}
\begin{center}
\begin{tabular}{c c c c  }
  \hline\hline
  $B$ &
  $E_c$&
  $\nu$ & $\chi^2_\text{red}$   %& $f(0)$
  \\
  \hline
  $\quad$0$\quad$ &
  $\quad$-3.259$\pm$0.005$\quad$ &  
 $\quad$2.73$\pm$0.02$\quad$ & $\quad$0.927$\quad$ %& 1.858
  \\
  1/1000 &
  -3.245$\pm$0.001 &
  3.43$\pm$0.08 & 0.843 %& 1.819
  \\
  1/500 &
  -3.242$\pm$0.002 &
  3.85$\pm$0.10  &  0.876 %& 1.852
  \\
  1/100 &
 -3.232$\pm$0.002 &
  4.47$\pm$0.15  &  0.890 %& 1.952
  \\
  \hline\hline
\end{tabular}
\end{center}
\caption{Table of the critical energy $E_c$, correlation length exponent 
$\nu$, and reduced chi square $\chi^2_{\text{red}}$ for different values of magnetic field. }
\end{table}

%by ....\YM{This figure is missing, we'll add more text once it is included.}
%{\YA The collapse of all the curves by a single length, $\xi\sim(E-E_c)^{-\nu}$, proves, as is usual for the transfer matrix method,  that the observation is not a finite size effect.}
%
%\YM{ YM: I think that the phase diagram in the $E-B$ plan should be a separate figure. I would prefer to switch the axes. I would also like to add more points to the curve $E_c(B)$. I thought that increasing $B$ should increase the range of the extended states. Are we sure that this is the opposite ? In particular, say we are on the localized side of the transition. How does $\xi$ depend on $B$ ? My recollection is that it was shown the $\xi$ increases with $B$, which does not seem to me to be consistent with the band of extended states becoming narrower.\\
%\ \\
%\underline{ Yshai}: I understand that the issue of $\xi(B)$ is resolved. I also think that the phase diagram should be
%displayed, and I continue my polishing job assuming that the phase diagram is included. }

\begin{figure}[b]
  \begin{center}% Requires \usepackage{graphicx}
  \includegraphics[width=8 cm]{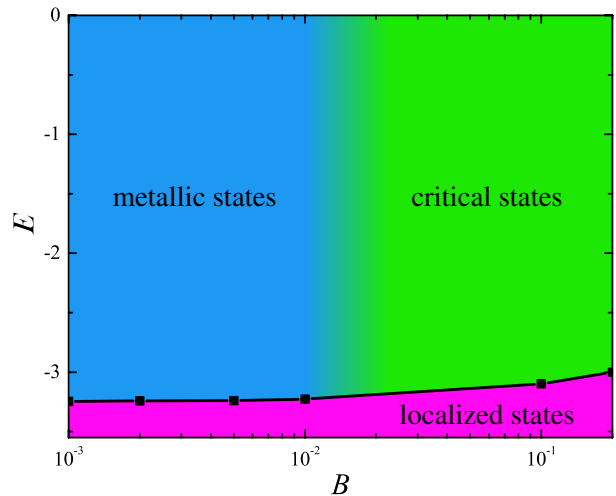}
  \end{center}
  \vspace{-0.2in}
  \caption{(color online). \footnotesize{Schematic phase diagram in the $E$-$B$
plane displaying the occurrence of three phases of localized states (pink),
metallic states (blue), and critical states  (green) \cite{note0}. See text for further details.}}
%\YMq{I don't like the bold line separating the extended and critical states, we don't claim a sharp transition. I suggest that the blue extended states will gradually change to the green extended states over some finite region around B=1/70.}}
  \label{fig3}
 \vspace{-0.2in}
\end{figure}

%\noindent
%\underline{Phase diagram}
In order to present a broader picture of the nature of states in the 2DSU model, we
combine the results of the present study with those of Ref.~\cite{Chen2}, where the existence of a band of critical states at strong magnetic
fields ($B \ge 1/5$) has been demonstrated. It is found that $E_c(B)$
is a slowly increasing function, and that somewhere around $B=1/70$ the Anderson MIT (discussed here) crosses over into a BKTT discussed previously \cite{Chen2} \cite{crossover}.
% For the moment we just display the schematic phase diagram in the case where this curve is simply a straight line $B=B_c \simeq 1/70$ where $B_c$ is referred to as a critical field.
%So far we have shown existence of mobility edges at a few values of the magnetic field, $E_c(B=0)<E_c(B=1/1000)<E_c(B=1/100)$.
%for $B=0$ and for Following the same procedure as a function of $B$, we can trace the critical curves in the $E-B$ plane.
The resulting phase diagram in the $E$-$B$ plane is depicted in Fig. \ref{fig3}. The emerging picture is that
 the band of extended states known to exist at $B=0$, persists for finite $B$, until strong enough magnetic field $B \simeq 1/70$ it crosses over (either sharply or smoothly) into a band of critical states as discussed in Ref. \cite{Chen2}.
\noindent
\underline{Summary:}
%A system composed of 2DEG on a lattice, subject to disorder, perpendicular
%magnetic field and random SOC (
%The 2DSU model, whose Hamiltonian is given in Eq.~(\ref{2DSU}), supports
%a BKTT from a band of localized states to a band of
%{\it critical states} at strong magnetic field  \cite{Chen2}.
Starting from the 2DSU model Hamiltonian (Eq.~\ref{2DSU}),
 we focus on the localization issue at the weak field regime, starting at $B=0$ where it is known to display  MIT for system with the symplectic symmetry.
% First, it is established that BKTT
%exists already above much lower, $B_c \approx 1/70$ flux per square, see
%Fig.~\ref{fig1}b.  Second,
Based on analyses of level statistics (Fig.~\ref{fig1}) and
localization length (Fig.~\ref{fig2}), it has been demonstrated that
a band of metallic states persists also for finite magnetic field
$0<B<B_c \simeq 1/70$. Combined with our previous results  \cite{Chen2},
we can suggest a schematic phase diagram in Fig. \ref{fig3}, that
elucidates the nature of localization in the $E$-$B$ plane
under the influence of spatially random spin-orbit potential.
% there is an ALT
%between a band of localized states and a band of EMS, see Fig.~\ref{fig1}a.
%These results are summarized in Fig.~\ref{fig1}c.
%The occurrence of EMS even for finite (albeit weak) magnetic field is at
%odd with the paradigm stating in 2DEG disordered system with unitary
%symmetry all states are localized.
Thus, the paradigm that all states in 2D disordered systems
with unitary symmetry are localized should be reviewed
when strong spin-orbit fluctuations are present.
In other words, in contrast to the prediction of the one-parameter scaling
theory of localization \cite{Abrahams}, localization in 2D disordered
systems is not unambiguously determined by its symmetry. This suggests that, similar to what happens in the quantum Hall regime (occurring at  strong magnetic field, without spin-orbit scattering), a second parameter
is required to describe the scaling of the dimensionless conductance. 
Obvious questions are how to introduce such a parameter, and how the renormalization-group
flow will look like in the presence of this additional parameter. Presently, the answers remain
a theoretical challenge.
 %From scaling theory point of view, our result calls for its extension  appropriate
%The point is that for  systems under weak magnetic field
%and strong random spin-orbit fluctuations.

%should be worked out, but so far it has not been established.

Remarkably, (and unlike the localization issue), level statistics is found to be
determined solely by symmetry, whether states are metallic or critical.
 As shown in Fig.~\ref{fig1}, for $B=0$, $P(s)$
follows the Wigner surmise for the GSE, while for $B=1/10100$, $P(s)$
follows the Wigner surmise for the GUE. Moreover, $P(s)$ obeys the
GUE statistics also for the band of critical states discussed in Ref. \cite{Chen2}.
This latter band is obtained following BKTT at strong magnetic field. In contrast,
for critical states around a mobility edge in a standard Anderson MIT,
a novel $P(s)$ statistics is suggested  \cite{newclass}. What we conclude here is
that $P(s)$ is the same for metallic and critical states and depends solely on symmetry.
% (see Fig.~\ref{fig2}).

%
%In addition to the study of localization within the transfer matrix
%techniques, we have attempted to get knowledge on the nature of quantum
%states and their spectra, through the study of level statistics
%and multi-fractal spectrum. Somewhat unexpectedly,
%the nearest-neighbor
%level spacing distribution $P(s)$ of states in the critical band
%is well described by the Wigner-Dyson distribution pertaining to GUE.
%Similar behaviour holds for the number variance of energy levels
%$\Sigma_2(N)$ of the critical states provided $\langle N \rangle<6$.
%For higher $\langle N \rangle$,  $\Sigma_2(N)$  displays system size
%dependence. Our results do not support the conjecture that level statistics
%of critical states follows a new universality class.
%It is also remarkable to note that for Finally, it is shown that the multi-fractal spectral function $f(\alpha)$
%is adequately described by analytic expressions derived from non-linear
%$\sigma$ model  \cite{conductance}.

\begin{acknowledgements}
This work is supported by NSFC of China grant (11374249) and Hong Kong RGC
grants (163011151 and 605413). The research of Y.A is partially supported
by grant 400/12 of the Israeli Science Foundation. Y. A acknowledges fruitful discussions with T. Ohtsuki. Y.M. acknowledges ISF grant 292/15.
\end{acknowledgements}


\begin{thebibliography}{99}
\bibitem{note1}As long as $W<W_c \simeq 6.3$, the results are similar.
The model does not support any extended states for $W>W_c$.
\bibitem{note3}The density of states is flat in the energy range $0.5\leq
E\leq 2.5$ so that the average of level spacing is a constant in the range.
One expects that the energy average and the ensemble average of level
spacing distribution function $P(s)$ are the same.
\bibitem{crossover}
Elucidating the nature of this crossover is beyond our scope. It requires the calculations of $\bar{\lambda}_M$
for many points in the $E$-$B$ plane, with the hope to establish a critical curve
separating the two bands of metallic and critical states.
\end{thebibliography}
\end{document}